\documentclass{ieeetj}

\usepackage{cite}
\usepackage{amsmath,amssymb,amsfonts,mathtools,bm}
\usepackage{amsthm}
\usepackage{algorithm,algorithmic}
\usepackage{graphicx}
\usepackage{booktabs,tabularx,array,multirow}
\usepackage{xcolor}
\usepackage{tikz}
\usetikzlibrary{arrows.meta,positioning,calc,decorations.pathreplacing,fit,shapes.geometric}
\usepackage{microtype}
\usepackage{stfloats}
\usepackage{needspace}
\usepackage{balance}
\usepackage{url}
\usepackage{hyperref}
\hypersetup{hidelinks}

\def\BibTeX{{\rm B\kern-.05em{\sc i\kern-.025em b}\kern-.08em
    T\kern-.1667em\lower.7ex\hbox{E}\kern-.125emX}}
\AtBeginDocument{\definecolor{tmlcncolor}{cmyk}{0.93,0.59,0.15,0.02}\definecolor{NavyBlue}{RGB}{0,86,125}}

\def\OJlogo{}
\def\seclogo{}
\def\authorrefmark#1{\ensuremath{^{\textbf{#1}}}}

\newtheorem{theorem}{Theorem}
\newtheorem{corollary}{Corollary}
\newtheorem{lemma}{Lemma}
\DeclareMathOperator{\diag}{diag}
\DeclareMathOperator{\tr}{tr}
\newcommand{\R}{\mathbb{R}}
\newcommand{\OO}{\mathrm{O}}

\newcommand{\LSD}{\operatorname{LSD}}

\begin{document}
\receiveddate{October 2026. This work is a preprint and was submitted to the IEEE Open Journal of Signal Processing; it has not yet been peer reviewed}
\reviseddate{}
\makeatletter\let\@revised\@empty\makeatother
\accepteddate{}
\publisheddate{}
\currentdate{}
\doiinfo{}

\markboth{IEEE Open Journal of Signal Processing}{Franchino and Prawda: Structured Pruning of Feedback Delay Networks}

\title{Structured Pruning of Feedback Delay Networks}

\author{Facundo Franchino\authorrefmark{1,2}, Karolina Prawda\authorrefmark{1,3}}
\affil{AudioLab, School of Physics, Engineering and Technology, University of York, York, United Kingdom}
\affil{Massachusetts Institute of Technology, Cambridge, MA, USA}
\affil{Acoustics Lab, Department of Information and Communications Engineering, Aalto University, Espoo, Finland}
\corresp{Corresponding author: Facundo Franchino (email: facundof@mit.edu).}

\begin{abstract}
Feedback delay networks (FDNs) are recursive filter structures used to synthesise artificial reverberation. Their feedback matrix governs both energy exchange between delay lines and the computation required at every sample, making its design central to balancing an FDN's acoustic performance and implementation efficiency. In this work, we develop an FDN-specific structured pruning method that connects these acoustic and computational concerns, allowing high-order dense orthogonal designs to be fitted to efficient implementations. The feedback matrix is replaced by stages of fixed permutations and small orthogonal blocks. We characterise connectivity growth with the network depth and derive exact conditional block updates for matrix fitting, followed by refinement against the reference FDN magnitude response. The resulting factors admit direct vectorised evaluation, with arithmetic cost and coefficient storage scaling linearly with network size for fixed block size and depth. On an Apple M2, the selected staged kernels achieve feedback-multiplication speedups over dense BLAS ranging from approximately threefold to over twentyfold across the tested network sizes, while substantially reducing coefficient storage. Objective metrics and listening tests indicate that the proposed staged approximations closely preserve the spectral and temporal character of their dense references. An analysis of processor utilisation and register reuse explains why execution time depends on more than the operation count. Finally, block size and depth provide practical control over response fidelity, storage, and real-time cost, allowing acoustic design to precede the choice of an efficient implementation.
\end{abstract}

\begin{IEEEkeywords}
Artificial reverberation, feedback delay networks, orthogonal matrix factorisation, real-time signal processing, differentiable signal processing, SIMD optimisation, structured sparsity.
\end{IEEEkeywords}

\maketitle

\section{Introduction}
\label{sec:introduction}

\IEEEPARstart{F}{eedback} Delay Networks (FDNs) remain a central class of algorithmic reverberators. Delay lines coupled by a feedback matrix can generate long, dense decays with modest computational resources~\cite{jot1991,stautner1982multichannel,valimaki2012fifty}. Although the delays and the input and output gains shape the FDN's response, the feedback matrix influences how readily energy mixes, how evenly the system modes are excited, and how much computation each sample requires \cite{jot1997efficient, menzer2010unitary}. This makes the feedback matrix a crucial design element for an FDN reverberator~\cite{jot1997efficient,schlecht2020fdntb}.

\begin{figure}[!t]
    \centering
    \resizebox{0.82\columnwidth}{!}{\begin{tikzpicture}[>=latex, font=\rmfamily]

    \def\R{3.8}
    \def\shift{2.2}
    \coordinate (C1) at (90:\shift);
    \coordinate (C2) at (210:\shift);
    \coordinate (C3) at (330:\shift);

    \draw[thick] (C1) circle (\R);
    \draw[thick] (C2) circle (\R);
    \draw[thick] (C3) circle (\R);

    \node[fill=white, font=\Large, align=center, inner sep=4pt] at (90:\shift+\R-1.5) {Colourless\\[0.3em] \normalsize\normalfont Metric: modal std $\sigma_\rho$};

    \node[fill=white, font=\Large, align=center, inner sep=4pt] at (210:\shift+\R-1.5) {Smooth\\[0.3em] \normalsize\normalfont Metric: mixing time};

    \node[fill=white, font=\Large, align=center, inner sep=4pt] at (330:\shift+\R-1.5) {Cheap\\[0.3em] \normalsize\normalfont Metric: FLOPs / SIMD};

    \node[font=\small, align=center] at (150:\shift+0.4) {Dense Matrices\\[0.1em] \footnotesize (Too slow)};

    \node[font=\small, align=center] at (30:\shift+0.4) {\\[0.1em] \footnotesize };

    \node[font=\small, align=center] at (270:\shift+0.4) {\\[0.1em] \footnotesize };

    \node[font=\Huge, align=center] at (0,0) {?};

\end{tikzpicture}}
    \caption{The ideal FDN triad combines even modal excitation, a smooth temporal response, and low deployment cost.}
    \label{fig:triad}
\end{figure}
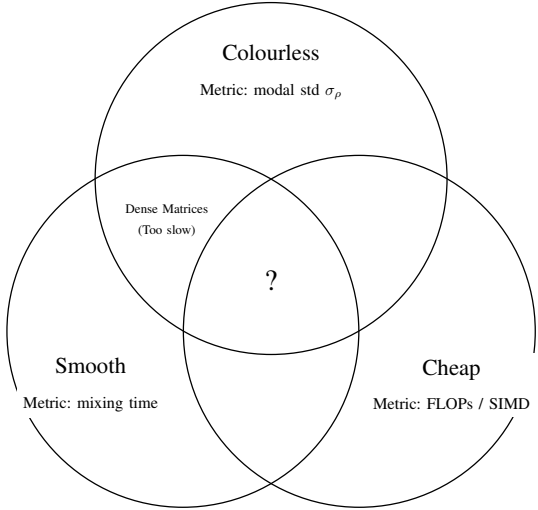

An FDN should produce spectrally colourless reverberation, apart from the intended frequency-dependent decay~\cite{dalsanto2023,dalsanto2025tiny,schlecht2017lossless}, with a smooth temporal response~\cite{schlecht2017echo} and low computational cost. However, achieving these three goals concurrently is non-trivial. Dense coupling affords design freedom at a high cost, while sparsity reduces the computation but can impede the exchange of energy within the system, leading to spectral colouration and slow temporal mixing. Figure~\ref{fig:triad} illustrates these competing aims together as the \textit{ideal FDN triad}.

Recent differentiable methods train dense feedback matrices against spectral and temporal objectives, using a matrix exponential to enforce orthogonality~\cite{dalsanto2023,santo2025,dalsanto2025tiny}. Such a matrix can meet the spectral and temporal aims of the \emph{ideal FDN triad}, but remains expensive to apply at every sample. To reduce this cost, one approach is to optimise a small FDN directly~\cite{dalsanto2025tiny}. At such low orders, however, smooth and colourless reverberation is harder to obtain and depends on careful parameter selection~\cite{dalsanto2023,dalsanto2025tiny}.

Naturally, efficient feedback matrices predate differentiable FDN design. Specific matrix families, such as Householder \cite{gerzon1976, householder1958triangularization}, Hadamard \cite{jot1997efficient, hadamard1893resolution} and circulant \cite{rocchesso1997} matrices offer savings compared to a random orthogonal design \cite{jot1997efficient}, although their spectral \cite{dalsanto2025tiny, heldmann2021} and temporal properties depend on the chosen structure \cite{prawda2020flexible}. Large sparse matrices have also been proposed \cite{menzer2010unitary, anderson2015flatter}.

Software tools also connect FDN design with implementation. FDNTB~\cite{schlecht2020fdntb} and its Python successor pyFDN~\cite{schlecht2026pyfdntoolbox} provide routines for network construction and analysis; pyFDN also integrates differentiable optimisation of such designs. ADAC compiles trained audio graphs to real-time FAUST processors while retaining their parameters and topology, making trained FDNs viable for deployment to a plethora of targets~\cite{franchino2026}. Compilation preserves the precise trained structure, however, so the cost of a dense feedback matrix follows it onto every target.

Rather than beginning with a small structure and optimising its response, this work starts with a dense feedback matrix that already has desirable acoustic properties and seeks to retain them in a cheap representation. The dense matrix is approximated by a product of sparse matrices, each describing fixed routing or independent small orthogonal blocks. These component matrices are the \emph{factors} of the staged representation. Thus, acoustic design precedes the choice of an efficient feedback implementation.

Two-channel rotations and recursive Kronecker constructions provide parametric families of orthogonal feedback matrices, designed directly in their structured form~\cite{gerzon1971synthetic1,menzer2010unitary,coppola2026fast}. Here, the task is instead to fit a specified dense reference within a staged family chosen for efficient execution. Classical Procrustes solutions give exact conditional block updates, while subsequent refinement adjusts all block parameters together to fit the complete FDN response. Furthermore, we provide specialised C++ kernels to evaluate the resulting stages directly.

\section{FDNs and Structured Feedback}
\label{sec:fdn}

\subsection{Feedback Delay Networks}

\begin{figure}[t]
    \centering
    \resizebox{\columnwidth}{!}{\definecolor{fdnblue}{rgb}{0.227,0.522,0.969}
\definecolor{fdnbluefill}{rgb}{0.820,0.902,0.992}
\begin{tikzpicture}[
    >=stealth,
    auto,
    delay/.style={draw, rectangle, minimum width=3cm, minimum height=0.6cm, line width=.55pt, fill=white},
    matrixA/.style={draw=fdnblue, rectangle, minimum width=3.5cm, minimum height=3.5cm, line width=1.5pt, fill=fdnbluefill, postaction={draw=black,line width=.35pt,fill=none}},
    sum/.style={draw, circle, inner sep=0pt, minimum size=0.35cm, line width=.55pt, fill=white},
    amp/.style={regular polygon, regular polygon sides=3, draw, line width=.55pt, minimum size=0.5cm, inner sep=0pt, shape border rotate=-90, fill=white},
    font=\large
]

\node[matrixA] (A) at (0.5, 3) {};
\ifdefined\fdnstaged
    \node[font=\Huge,inner sep=0pt] at (.5,4.15) {$\mathbf S$};
    \node[draw,fill=white,line width=.55pt,minimum width=1.05cm,minimum height=.8cm,inner sep=2pt,font=\Large] (first) at (1.55,3.05) {$\Phi_0$};
    \node[inner sep=0pt,font=\Large] (dots) at (.5,3.05) {$\cdots$};
    \node[draw,fill=white,line width=.55pt,minimum width=1.05cm,minimum height=.8cm,inner sep=2pt,font=\Large] (last) at (-.55,3.05) {$\Phi_{L-1}$};
    \draw[->,line width=.55pt] (first.west)--(dots.east);
    \draw[->,line width=.55pt] (dots.west)--(last.east);
    \node[font=\Large,inner sep=0pt] at (.5,1.95) {$\Phi_\ell=\mathbf B_\ell\mathbf P_\ell$};
\else
    \node[font=\Huge,inner sep=0pt,outer sep=0pt,anchor=center] at (A.center) {$\bm A$};
\fi

\def\xInDrop{-5.8}
\def\xAmpB{-4.8}
\def\xDelay{0.5}
\def\xTap{3.4}
\def\xAmpC{5.8}
\def\xSumOut{6.8}
\def\yDirect{-4.8}

\node (xn) at (-6.8, 0) {$x(n)$};
\draw[line width=.55pt] (xn) -- (\xInDrop, 0);
\draw[line width=.55pt] (\xInDrop, 0) -- (\xInDrop, \yDirect);

\foreach \i/\y/\ay/\xDrop in {1/0/1.5/-2.8, 2/-1.2/2.5/-3.2, 3/-2.4/3.5/-3.6, 4/-3.6/4.5/-4.0} {
    \node[amp] (b\i) at (\xAmpB, \y) {};
    \node[above=0.05cm of b\i] {$b_\i$};
    \node[sum] (sumin\i) at (\xDrop, \y) {$+$};
    \node[delay] (d\i) at (\xDelay, \y) {$z^{-m_\i}$};
    \node[amp] (c\i) at (\xAmpC, \y) {};
    \node[above=0.05cm of c\i] {$c_\i$};
    \node[sum] (sumout\i) at (\xSumOut, \y) {$+$};
    \coordinate (tap\i) at (\xTap, \y);

    \draw[->, line width=.55pt] (\xInDrop, \y) -- (b\i);
    \draw[->, line width=.55pt] (b\i) -- (sumin\i);
    \draw[->, line width=.55pt] (sumin\i) -- (d\i.west |- 0,\y);
    \draw[line width=.55pt] (d\i.east |- 0,\y) -- (tap\i);
    \node[anchor=north,inner sep=1pt] at (2.75,{\y-.12}) {$s_\i(n)$};
    \draw[->, line width=.55pt] (tap\i) -- (c\i);
    \draw[->, line width=.55pt] (c\i) -- (sumout\i);

    \draw[->, line width=.55pt] (A.west |- 0,\ay) -| (sumin\i.north);
}

\foreach \i/\y/\ay/\xRise in {1/0/1.5/3.8, 2/-1.2/2.5/4.2, 3/-2.4/3.5/4.6, 4/-3.6/4.5/5.0} {
    \draw[->, line width=.55pt] (tap\i) -- (\xRise, \y) |- (A.east |- 0,\ay);
}

\node[amp] (d) at (0.5, \yDirect) {};
\node[above=0.05cm of d] {$d$};
\draw[->, line width=.55pt] (\xInDrop, \yDirect) -- (d);
\draw[->, line width=.55pt] (d) -| (sumout4.south);

\draw[->, line width=.55pt] (sumout4.north) -- (sumout3.south);
\draw[->, line width=.55pt] (sumout3.north) -- (sumout2.south);
\draw[->, line width=.55pt] (sumout2.north) -- (sumout1.south);
\draw[->, line width=.55pt] (sumout1.east) -- (7.8, 0) node[right] {$y(n)$};

\end{tikzpicture}}
    \caption{Lossless four-line FDN ($\bm\Gamma=\mathbf I$). Projection replaces the feedback matrix $\mathbf A$, highlighted in blue, with the staged matrix $\mathbf S$. Delays and gains remain fixed; the evaluated systems also retain their delay-proportional attenuation.}
    \label{fig:fdn_diagram}
\end{figure}
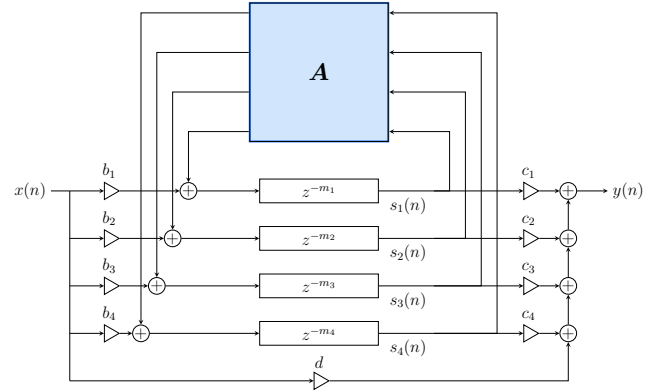

A single-input, single-output (SISO) FDN consists of $N$ delay lines with positive integer lengths $m_i$ in samples, an $N\times N$ feedback matrix $\mathbf A$, input and output gain vectors $\mathbf b,\mathbf c\in\R^N$, and an optional direct gain $d$~\cite{jot1991, dalsanto2023,santo2025,dalsanto2025tiny}. Let $\bm\Gamma=\diag(\gamma^{m_1},\ldots,\gamma^{m_N})$ give delay-proportional attenuation, with $0<\gamma\leq1$. The FDN transfer function is then~\cite{schlecht2017lossless}
\begin{equation}
\begin{aligned}
H(z)&=\mathbf c^{\top}\!\left[\mathbf D_{\mathbf m}(z)-\mathbf A\bm\Gamma\right]^{-1}\mathbf b+d,\\
\mathbf D_{\mathbf m}(z)&=\diag(z^{m_1},\ldots,z^{m_N}).
\end{aligned}
\label{eq:fdn_transfer}
\end{equation}
Figure~\ref{fig:fdn_diagram} illustrates the corresponding signal flow. The delays $m_i$ establish the temporal scale and modal resolution of the resulting reverberation \cite{schlecht2017echo}, while $\mathbf A$ redistributes energy between the delay states, marked in Fig.~\ref{fig:fdn_diagram} by $s_i$, at every recursion \cite{jot1991}.


We use a real orthogonal lossless prototype, so that the feedback matrix preserves energy and decay is controlled by $\bm\Gamma$~\cite{schlecht2017lossless}. The matrix satisfies
\begin{equation}
\mathbf A^{\top}\mathbf A=\mathbf I
\label{eq:orthogonal}
\end{equation}
which allows mixing and decay to be analysed separately~\cite{gerzon1976,schlecht2017lossless}. The lossless case has $\gamma=1$ and $\bm\Gamma=\mathbf I$. In the implementation used in this study, attenuation precedes mixing, leading to the feedback operator $\mathbf A\bm\Gamma$.

\subsection{A Pruning Paradigm in FDNs}

Merely reducing the number of delay lines lowers computational cost, but can introduce spectral colouration \cite{dalsanto2025tiny} and audibly sparse output \cite{schlecht2017echo}. Here, a larger dense FDN with the desired acoustic behaviour is taken as the starting point, and we introduce an FDN-specific adaptation of structured pruning to reduce its implementation cost while retaining its delay lines.

Neural-network pruning removes individual weights~\cite{frankle2019lottery} or structured groups of parameters, including in real-time audio effects~\cite{clarke2025}. Here, pruning means projection onto a constrained operator family. Given a dense orthogonal $\mathbf A$, the problem is to find a structured orthogonal $\mathbf S$ that approximates the reverberator while admitting a low-cost implementation. Sparsity is imposed on the factors, so their product $\mathbf S$ need not be sparse. Evaluation uses the stored blocks and fixed routing directly, which means entries excluded by the routing require neither coefficient storage nor multiplication~\cite{duff2017}. Furthermore, the delay lengths, gains, and attenuation are fixed, so every measured change follows from the feedback projection.

\begin{figure}[t]
    \centering
    \begin{tikzpicture}[x=1cm,y=1cm,font=\small,line width=.35pt,
    vertex/.style={circle,fill=black,inner sep=0pt,minimum size=2.5pt},
    >={Stealth[length=2.8pt,width=2.2pt]}]
    \node[inner sep=0pt] at (0,0) {$\mathbf M=\begin{bmatrix}
        \times & & \times & \\
        \times & & \times & \\
        & \times & & \times \\
        & \times & & \times
    \end{bmatrix}$};
    \node[anchor=south] at (2.5,1.02) {Input $j$};
    \node[anchor=south] at (4.35,1.02) {Output $i$};
    \foreach \i in {0,...,3} {
        \node[vertex] (x\i) at (2.5,{.81-.54*\i}) {};
        \node[vertex] (y\i) at (4.35,{.81-.54*\i}) {};
        \node[anchor=east,inner sep=3pt] at (x\i.west) {$\i$};
        \node[anchor=west,inner sep=3pt] at (y\i.east) {$\i$};
    }
    \foreach \j/\i in {0/0,2/0,0/1,2/1,1/2,3/2,1/3,3/3} {
        \draw[->,shorten <=1pt,shorten >=1pt] (x\j) -- (y\i);
    }
\end{tikzpicture}
    \caption{Matrix pattern (left) and bipartite graph (right) of pairwise routing at $N=4$. Inputs $(0,2)$ feed outputs $(0,1)$; inputs $(1,3)$ feed outputs $(2,3)$. Crosses mark allowed coefficient positions, not numerical values. Blanks mark coefficients fixed at zero by the routing. Each arrow represents an allowed entry $M_{ij}$ from input $j$ to output $i$.}
    \label{fig:sparsity_graph}
\end{figure}
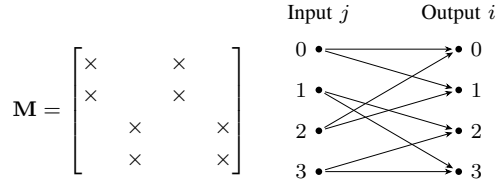

Sparsity has both an algebraic and a graph interpretation~\cite{duff2017}. In the signal-flow convention used here (\emph{cf.} Fig.~\ref{fig:fdn_diagram}) each delay line is a node, and a non-zero entry $A_{ij}$ couples state $j$ to state $i$, as in $y_i=\sum_j A_{ij}x_j$. Separating input and output nodes gives the bipartite representation in Fig.~\ref{fig:sparsity_graph}. Blank positions in the matrix remain zero for every choice of rotation angles because the routing provides no connection there. Crosses mark the permitted connections, whose coefficients depend on those angles and may also be zero. For example, a rotation through zero passes its inputs through without mixing them. Changing the angle can mix the same inputs without changing the routing. The graph therefore describes the connections available to the fit, rather than the values of the fitted coefficients.

\begin{figure}[t]
    \centering
    \resizebox{0.98\columnwidth}{!}{\begin{tikzpicture}[font=\rmfamily\small,x=1cm,y=1cm,
 dot/.style={circle,fill=black,inner sep=1.05pt},
 heading/.style={anchor=west,font=\rmfamily\small}]
\node[heading] at (0,0) {(a) Disconnected pairs};
\foreach \i/\j in {0/1,2/3,4/5,6/7}
 \draw[black,line width=.35pt] (.3+.95*\i,-.48) to[bend left=45] (.3+.95*\j,-.48);
\foreach \i in {0,...,7} \node[dot] at (.3+.95*\i,-.48) {};

\node[heading] at (0,-1.18) {(b) Adjacent couplings};
\foreach \i/\j in {0/1,2/3,4/5,6/7}
 \draw[black,line width=.35pt] (.3+.95*\i,-1.68) to[bend left=45] (.3+.95*\j,-1.68);
\foreach \i/\j in {1/2,3/4,5/6}
 \draw[black,densely dashed,line width=.35pt] (.3+.95*\i,-1.68) to[bend right=45] (.3+.95*\j,-1.68);
\foreach \i in {0,...,7} \node[dot] at (.3+.95*\i,-1.68) {};

\node[heading] at (0,-2.50) {(c) Staged paths from one input};
\foreach \l in {0,1,2} {
 \pgfmathtruncatemacro{\count}{pow(2,\l)}
 \pgfmathtruncatemacro{\last}{\count-1}
 \foreach \p in {0,...,\last} {
  \pgfmathsetmacro{\yp}{-4.34+(\count-1-2*\p)*1.28/\count}
  \foreach \j in {0,1} {
   \pgfmathtruncatemacro{\o}{2*\p+\j}
   \pgfmathsetmacro{\yo}{-4.34+(2*\count-1-2*\o)*1.28/(2*\count)}
   \draw[black,line width=.35pt] (.35+2.1*\l,\yp) -- (2.45+2.1*\l,\yo);
  }
 }
}
\foreach \l in {0,...,3} {
 \pgfmathtruncatemacro{\count}{pow(2,\l)}
 \pgfmathtruncatemacro{\last}{\count-1}
 \foreach \p in {0,...,\last} {
  \pgfmathsetmacro{\yp}{-4.34+(\count-1-2*\p)*1.28/\count}
  \node[circle,fill=black,inner sep=1.05pt] at (.35+2.1*\l,\yp) {};
 }
}
\node[font=\rmfamily\scriptsize] at (.35,-2.98) {Input};
\foreach \l in {1,2,3} \node[font=\rmfamily\scriptsize] at (.35+2.1*\l,-2.98) {Layer \l};
\foreach \l/\count in {1/2,2/4,3/8}
 \node[font=\rmfamily\scriptsize,text=black] at (.35+2.1*\l,-5.78) {\count\ outputs};
\end{tikzpicture}}
    \caption{Connectivity at $N=8$. (a) Independent rotations leave disconnected pairs. (b) Alternating adjacent couplings form a chain. (c) Fixed routing expands the paths from one input to two, four, then all eight outputs. Only that input's paths are drawn.}
    \label{fig:coupling}
\end{figure}
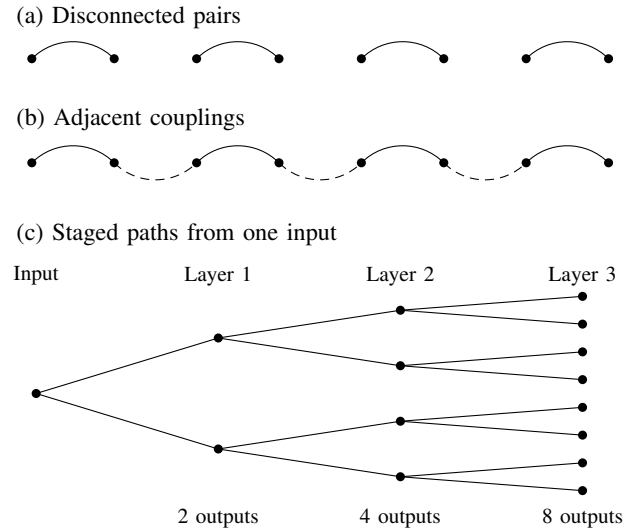

The benefits of pruning depend on connectivity, since the feedback couplings must provide paths through which energy can spread between delay lines. Figure~\ref{fig:coupling} illustrates examples of different couplings. Independent $2\times2$ blocks in Fig.~\ref{fig:coupling}(a) leave disconnected channel pairs. In Fig.~\ref{fig:coupling}(b), the adjacent rotations underlying Cantero--Moral--Vel\'{a}zquez matrices~\cite{cantero2003cmv,simon2005} connect the channels, but energy still has to cross a chain, leading to slow build-up of echoes in the produced reverberation. This illustrates why linear arithmetic cost alone does not guarantee rapid mixing.

This study follows the approach illustrated in Fig.~\ref{fig:coupling}(c), in which interleaving local rotations with a fixed permutation allows distant channels to meet at the next layer, widening the exchange of energy at a controlled arithmetic cost. The permutation preserves orthogonality and does not add multiplications, while its fixed routing can be absorbed into addressing or register selection.

\section{The Block-Staged Orthogonal Family}\label{sec:family}

The feedback operator is applied as a sequence of routing and mixing stages. Each stage gathers the current state into disjoint groups, mixes each group with a small orthogonal block, and passes the result to the next stage. The same construction accommodates pairwise rotations and larger blocks.

\subsection{Staged Approximation}

The proposed staged approximation is specified by choosing a block size $b$ that divides the number of channels $N$, and a depth (number of layers) $L$. Each layer contains $Q=N/b$ independent $b \times b$ blocks.

\begin{figure}[t!]
    \centering
    \resizebox{\columnwidth}{!}{\input{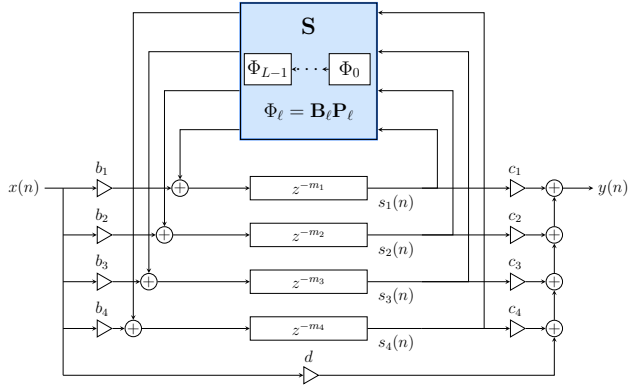}}
    \caption{The FDN of Fig.~\ref{fig:fdn_diagram} with staged feedback. Inside the highlighted block, $\Phi_0$ is applied first, followed by the remaining stages from right to left, along the feedback flow. Each stage gathers with $\mathbf P_\ell$ and mixes with $\mathbf B_\ell$. The product $\mathbf S$ is never formed in the audio path; the rest of the FDN is unchanged.}
    \label{fig:factorisation}
\end{figure}

$\mathbf W_{\ell k}\in\OO(b)$ is an orthogonal matrix that mixes selected entries of the input vector $\mathbf{x}$ for block $k$ in layer $\ell$, where $\OO(b)=\{\mathbf W:\mathbf W^\top\mathbf W=\mathbf I_b\}$ denotes the real orthogonal group. The indices start at zero. The indexing function (gather) $\pi_\ell(k,j)$ selects entry $j$ for block $k$ and must enumerate all $N$ channels exactly once. Each layer gathers these inputs and writes each block's $b$ outputs contiguously according to
\begin{equation}
[\Phi_\ell(\mathbf x)]_{bk+r}
 =\sum_{j=0}^{b-1}[\mathbf W_{\ell k}]_{rj}
 x_{\pi_\ell(k,j)},
\quad \substack{0\leq k<Q\\0\leq r<b}.
\label{eq:gather_layer}
\end{equation}
Thus, $j$ indexes a block input and $r$ its output. Equivalently, $\Phi_\ell=\mathbf B_\ell\mathbf P_\ell$, where $\mathbf B_\ell=\diag(\mathbf W_{\ell0},\ldots,\mathbf W_{\ell,Q-1})$ and $[\mathbf P_\ell\mathbf x]_{bk+j}=x_{\pi_\ell(k,j)}$. Starting with $\mathbf x^{(0)}=\mathbf x$, we apply $\mathbf x^{(\ell+1)}=\Phi_\ell(\mathbf x^{(\ell)})$ for $\ell=0,\ldots,L-1$. The complete operator is
\begin{equation}
\mathbf S\mathbf x=(\Phi_{L-1}\circ\cdots\circ\Phi_0)(\mathbf x)
                 =\mathbf x^{(L)}.
\label{eq:staged_apply}
\end{equation}
Here, $\circ$ denotes composition, with the rightmost map applied first. In matrix form, $\mathbf S=(\mathbf B_{L-1}\mathbf P_{L-1})\cdots(\mathbf B_0\mathbf P_0)$; the permutation matrices $\mathbf P_\ell$ are fixed, while the blocks of $\mathbf B_\ell$ are fitted. Figure~\ref{fig:factorisation} places these stages in the feedback path of the FDN in Fig.~\ref{fig:fdn_diagram}. Only the mixing operator changes: the delay lines, gains, and attenuation retain the values chosen for the acoustic reference.

\Needspace*{8\baselineskip}
\begin{theorem}[Orthogonality by construction]
For any fixed bijective gathers and any choice of blocks $\mathbf W_{\ell k}\in\OO(b)$, the operator $\mathbf S$ in~\eqref{eq:staged_apply} is orthogonal.
\end{theorem}
\Needspace*{3\baselineskip}
\begin{proof}
Each layer is the composition of a permutation and a block-diagonal orthogonal map. Because the product of orthogonal matrices remains orthogonal, the product of all layers, $\mathbf{S}$, is also orthogonal.
\end{proof}

For $b=2$, each block is a plane (Givens) rotation~\cite{golub2013,givens1958rotations}
\begin{equation}
\mathbf R(\theta)=
\begin{bmatrix}\cos\theta&-\sin\theta\\[1pt]\sin\theta&\cos\theta\end{bmatrix},
\label{eq:rotation}
\end{equation}
where $\theta$ is periodic modulo $2\pi$. The layer input $\mathbf x$  is split into its first and last $N/2$ entries. Then, block $k$ gathers $x_k$ and $x_{k+N/2}$, rotates the pair, and writes the results to positions $2k$ and $2k+1$. This contiguous interleaving implements the perfect shuffle~\cite{diaconis1983}.

For $b=4$, block $k$ reads the four inputs $k,k+Q,k+2Q,k+3Q$ and writes four contiguous outputs, giving the radix-4 Pease constant-geometry dataflow~\cite{pease1968}. Under these constant-geometry schedules, the structural support of an output (the set of inputs that can potentially influence it) spreads by at most a factor of $b$ per layer. Hence, $b^L\geq N$ is necessary for every output to be structurally dependent on every input. For the power-of-two sizes considered here, this bound is attained by the structural dependency graph, although it does not imply that every numerical entry is nonzero or that the FDN is already acoustically mixed.

\begin{lemma}[Structural spread]
Let $N=2^a$, $b=2^c$, and $1\leq c\leq a$, with gather $\pi(k,j)=k+jN/b$. Treat every block entry as an allowed coupling. After $L$ layers each output has structural paths from exactly $\min(b^L,N)$ inputs.
\end{lemma}

\Needspace*{3\baselineskip}
\begin{proof}
Tracing one layer backwards from output $o$ gives inputs $\lfloor o/b\rfloor+jN/b$, $0\leq j<b$. In binary, this deletes the lowest $c$ bits and inserts any $c$ bits at the top. After $L$ steps, $\min(cL,a)$ bits can be chosen freely and the others are fixed by $o$. There are therefore $2^{\min(cL,a)}=\min(b^L,N)$ reachable inputs. Particular block values may cancel paths, so the claim is structural.

\end{proof}

Counting each scalar multiplication or addition as one operation, a pairwise rotation costs four multiplications and two additions. Applying all $L$ layers therefore costs $3NL$ operations. A stored $4\times4$ block costs 16 multiplications and 12 additions, giving $7NL$ operations. The separate block-size benchmark counts the pairwise structure at its exact $3NL$ and uses the nominal multiply--accumulate convention $2bNL$ for the larger blocks, which includes the initial accumulation in the count.

Runtime coefficient storage is $NL$ floats for pairwise $(\cos\theta,\sin\theta)$ values and $4NL$ floats for radix-4 blocks. These counts are linear in $N$ for fixed $L$, whereas a dense feedback matrix stores $N^2$ coefficients and requires approximately $2N^2$ scalar operations per sample. Radix-4 reaches full structural spread in $\lceil\log_4N\rceil$ layers, compared with $\log_2N$ pairwise layers. For fixed block size, choosing the minimum depth for full structural spread therefore gives $O(N\log N)$ arithmetic and coefficient storage.

\subsection{Exact Conditional Projection}

The matrix projection seeks $\mathbf{S}$ that minimises the Frobenius error $\|\mathbf A-\mathbf S\|_F$. Since the squared Frobenius norm is the sum of squared entrywise differences, and both $\mathbf{A}$ and $\mathbf{S}$ are orthogonal with $\|\mathbf{A}\|_F^2 = \|\mathbf{S}\|_F^2 = N$, we arrive at
\begin{equation}
\|\mathbf A-\mathbf S\|_F^2
 =2N-2\tr(\mathbf A^\top\mathbf S).
\label{eq:frob_trace}
\end{equation}
Thus, minimising the Frobenius error is equivalent to maximising the trace $\tr(\mathbf A^\top\mathbf S)$.

To update layer $\ell$, all other layers are kept fixed and $\mathbf S=\mathbf M_{>}\mathbf B_\ell\mathbf P_\ell\mathbf M_{<}$, where $\mathbf M_{<}$ and $\mathbf M_{>}$ denote the earlier ($0, \ldots, \ell-1$) and later ($\ell+1, \ldots,  L-1$) layers, respectively, with an empty product yielding $\mathbf I_N$. Applying the cyclic invariance of the trace gives
\begin{equation}
\mathbf G_\ell=\mathbf P_\ell\mathbf M_{<}\mathbf A^\top\mathbf M_{>},
\qquad
\tr(\mathbf A^\top\mathbf S)=\sum_{k=0}^{Q-1}
\tr(\mathbf G_{\ell k}\mathbf W_{\ell k}),
\label{eq:environment}
\end{equation}
where $\mathbf G_{\ell k}$ is the $k$th diagonal block of $\mathbf G_\ell$. Since the summation decomposes into independent terms, each of them can be maximised independently as a local orthogonal Procrustes problem~\cite{schonemann1966,golub2013}.

\begin{theorem}[Exact block update]
Let $\mathbf G_{\ell k}=\mathbf U\bm\Sigma\mathbf V^\top$ be a singular value decomposition (SVD) with non-increasing singular values. Fix the determinant sign of the block $\delta\in\{-1,1\}$, and define the sign-correction matrix $\mathbf D_\delta=\diag(1,\ldots,1,\delta\det(\mathbf V\mathbf U^\top))$. A global maximiser of $\tr(\mathbf G_{\ell k}\mathbf W_{\ell k})$ over $\{\mathbf W\in\OO(b):\det\mathbf W=\delta\}$ is given by
\begin{equation}
\mathbf W_{\ell k}^{\star}=\mathbf V\mathbf D_\delta\mathbf U^\top.
\label{eq:procrustes}
\end{equation}
If both orientation components are admitted (i.e., $\mathbf W\in\OO(b)$ without a determinant restriction), setting $\delta=\det(\mathbf V\mathbf U^\top)$ yields $\mathbf D_\delta=\mathbf I$.
\end{theorem}
\Needspace*{3\baselineskip}
\begin{proof}
Substitute $\mathbf W=\mathbf V\mathbf Q\mathbf U^\top$ where $\mathbf Q=\mathbf V^\top\mathbf W\mathbf U \in \OO(b)$. The trace simplifies to $\tr(\mathbf G_{\ell k}\mathbf W_{\ell k}) = \tr(\mathbf{\Sigma Q}) =\sum_i\sigma_iQ_{ii}$ with $\det(\mathbf{Q})=\delta\det(\mathbf V\mathbf U^\top) \equiv s$. For $s=1$, the diagonal entries of $\mathbf{Q}$ satisfy $Q_{ii}\leq1$ and the trace is bounded above by $\sum_i\sigma_i$, which is attained when $\mathbf Q=\mathbf I$. For $s=-1$, a real orthogonal matrix must have at least one eigenvalue equal to $-1$, implying $\tr\mathbf Q\leq b-2$. Rearranging the terms yields the bound
$\sum_i\sigma_i(1-Q_{ii})\geq\sigma_b(b-\tr\mathbf Q)\geq2\sigma_b$, which is attained when $\mathbf Q=\diag(1,\ldots,1,-1)$. In both cases $\mathbf{Q} = \mathbf{D}_\delta$, thus $\mathbf W=\mathbf V\mathbf D_\delta\mathbf U^\top$ is optimal, although it need not be unique.

\end{proof}

\Needspace*{9\baselineskip}
\begin{corollary}[Pairwise atan2 update]
For $b=2$ and $\mathbf W=\mathbf R(\theta)$,
$\tr(\mathbf G\mathbf R(\theta))=(G_{00}+G_{11})\cos\theta+(G_{01}-G_{10})\sin\theta$, hence
\begin{equation}
\theta^\star=\operatorname{atan2}(G_{01}-G_{10},\;G_{00}+G_{11}).
\label{eq:atan2}
\end{equation}
Here $\operatorname{atan2}(y,x)$ is the angle of the vector $(x,y)$. If both arguments vanish, every angle is optimal.
\end{corollary}
The pairwise implementation ($b=2$) enforces $\delta=+1$ and uses the closed-form angle update in~\eqref{eq:atan2}, whereas the radix-4 implementation ($b=4$) admits both determinant signs ($\delta=\pm1$). Within each layer $\ell$, all blocks $k$ can be updated independently. In exact arithmetic, each update can only increase or preserve the trace. Since $\tr(\mathbf A^\top\mathbf S)\leq N$, forward and backward sweeps produce convergent objective values. This establishes neither convergence of the block matrices nor a global minimum over all layers together.

\subsection{Joint Refinement in Local Coordinates}

The conditional updates optimise one layer while the others remain fixed, with independent block problems within that layer. Joint refinement instead varies every block in every layer in the same optimisation. To retain orthogonality throughout this simultaneous update, we express each block in unconstrained coordinates. For $b=2$, the rotation angle $\theta$ suffices. For larger blocks, a skew-symmetric matrix $\mathbf K^\top=-\mathbf K$ parametrises a Cayley transform around the fitted block $\mathbf W_0$~\cite{cayley1846,wen2013feasible},
\begin{equation}
\mathbf W(\mathbf K)=\mathbf C(\mathbf K)\mathbf W_0,
\qquad
\mathbf C(\mathbf K)=(\mathbf I+\mathbf K)^{-1}(\mathbf I-\mathbf K).
\label{eq:cayley}
\end{equation}
Here, only the $b(b-1)/2$ entries of $\mathbf{K}$ above the diagonal are free parameters; the remaining entries are fixed by skew symmetry. Since the eigenvalues of $\mathbf K$ are purely imaginary, $\mathbf I+\mathbf K$ is guaranteed to be invertible. Moreover, the Cayley transform $\mathbf C(\mathbf K)$ is proper orthogonal, i.e., $\mathbf C^\top\mathbf C=\mathbf I$ and $\det\mathbf C=1$, ensuring the preservation of both the orthogonality and the determinant sign of $\mathbf W_0$.

\emph{Lie-group viewpoint.} The orthogonal matrices form the Lie group $\OO(b)$. Its tangent space at the identity, the Lie algebra $\mathfrak{so}(b)$, consists of skew-symmetric matrices: these describe infinitesimal rotations~\cite{edelman1998}. The Cayley transform turns such coordinates into a finite rotation $\mathbf C(\mathbf K)$, which moves the fitted block $\mathbf W_0$ while preserving its determinant sign. Thus, optimisation takes place in unconstrained coordinates while every evaluated block remains orthogonal. This chart omits relative rotations $\mathbf W\mathbf W_0^\top$ with eigenvalue $-1$; it is not a global parametrisation of the group.

\section{Dense-to-Staged Projection}
\label{sec:projection}

\begin{figure*}[t]
    \centering
    \resizebox{0.77\textwidth}{!}{\begin{tikzpicture}[>=Latex,font=\rmfamily\normalsize,
 block/.style={draw=black,line width=.4pt,minimum height=1.35cm,align=center},
 flow/.style={->,line width=.4pt}]
\node (input) at (0,0) {$\mathbf A$};
\node[block,minimum width=3.5cm] (fit) at (2.6,0)
 {Matrix fitting\\[5pt]$\displaystyle\min_{\boldsymbol\theta}\;\|\mathbf A-\mathbf S(\boldsymbol\theta)\|_F^2$};
\node[block,minimum width=2.15cm] (fdn) at (7.2,0)
 {Staged FDN\\[5pt]$\mathbf S(\boldsymbol\theta)$};
\node[draw=black,line width=.4pt,minimum width=1.6cm,minimum height=.75cm] (loss) at (10.9,0)
 {$\mathcal L_{\mathrm{LSD}}$};
\draw[flow] (input) -- (fit);
\draw[flow] (fit) -- node[above] {$\boldsymbol\theta_0$} (fdn);
\draw[flow] (fdn) -- node[above] {$H_{\mathbf S}(\omega_k)$} (loss);
\node[align=center] (target) at (10.9,1.12) {Dense target\\[2pt]$H_{\mathbf A}(\omega_k)$};
\draw[flow] (target) -- (loss);
\draw[->,dashed,line width=.5pt,tmlcncolor] (loss.east) -- ++(.55,0)
 |- (7.2,2.05) node[pos=.65,above,text=black] {$\nabla_{\boldsymbol\theta}\mathcal L_{\mathrm{LSD}}$}
 -- (fdn.north);
\node[anchor=south] at (9.0,2.7) {Response refinement};
\draw[->,densely dotted,line width=.4pt] (fdn.south) -- ++(0,-.7)
 node[below] {$\boldsymbol\theta_{\mathrm{final}}$};
\end{tikzpicture}}
    \caption{Two-phase projection. Matrix fitting first approximates the entries of $\mathbf A$ and supplies the initial parameters $\boldsymbol\theta_0$. Response refinement then adjusts these parameters to match the staged FDN response $H_{\mathbf S}$ to the dense target $H_{\mathbf A}$ through the log-spectral loss $\mathcal L_{\mathrm{LSD}}$. The dashed blue path carries the gradient; the dotted arrow gives the final parameters. Delays, gains, and attenuation remain fixed.}
    \label{fig:pipeline}
\end{figure*}
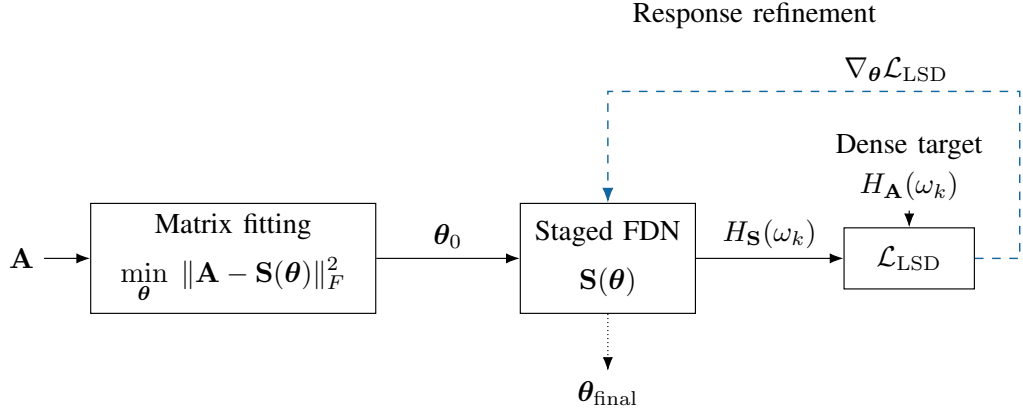

The dense-to-staged projection takes a dense orthogonal feedback matrix $\mathbf{A}$ and produces a set of parameters $\boldsymbol\theta$ such that the staged matrix $\mathbf S(\boldsymbol\theta)$ approximates the acoustic response of the dense FDN. Figure~\ref{fig:pipeline} shows the two phases, matrix fitting and response refinement; Algorithm~\ref{alg:projection} summarises their execution.

The matrix-fitting phase minimises the Frobenius distance between $\mathbf A$ and $\mathbf S$ by maximising $\tr(\mathbf A^\top\mathbf S)$. It uses 15--20 initialisations, comprising one greedy outside-in fit and random starts for the remaining candidates. Each candidate undergoes forward and backward sweeps using~\eqref{eq:procrustes}, or~\eqref{eq:atan2} for rotation pairs. The best trace value selects the initialisation for joint L-BFGS refinement, using angles for $b=2$ and Cayley coordinates for $b=4$. The pairwise implementation also tries perturbed L-BFGS restarts. Joint refinement allows coordinated changes across layers while remaining a local optimisation.

\begin{algorithm}[t]
\caption{Block-staged projection}
\label{alg:projection}
\begin{algorithmic}[1]
\REQUIRE Dense orthogonal $\mathbf A$, block size $b$, layers $L$, gathers $\pi_\ell$; fixed delays, gains, and attenuation
\STATE Initialise blocks greedily; add random restarts
\FOR{each restart}
  \REPEAT
    \FOR{$\ell=0,\ldots,L-1,L-1,\ldots,0$}
      \STATE Form $\mathbf G_\ell$ using~\eqref{eq:environment}
      \STATE Update all blocks by~\eqref{eq:procrustes} or~\eqref{eq:atan2}
    \ENDFOR
  \UNTIL{trace gain is below tolerance or the sweep limit is reached}
\ENDFOR
\STATE Keep the best restart; refine all block coordinates jointly
\STATE Refine against the FDN response using~\eqref{eq:lsd}; rescore the final blocks
\RETURN Selected orthogonal blocks and fixed routing
\end{algorithmic}
\end{algorithm}

A small matrix error need not give a small response error in a recursive system. Acoustic refinement therefore adjusts $\boldsymbol\theta$ against the FDN frequency response.

Define $\mathbf R_{\mathbf A}(z)=[\mathbf D_{\mathbf m}(z)-\mathbf A\bm\Gamma]^{-1}$ and define $\mathbf R_{\mathbf S}(z)$ analogously. These resolvents describe the recursive response for the dense and staged feedback matrices. The identity $X^{-1}-Y^{-1}=X^{-1}(Y-X)Y^{-1}$ gives
\begin{equation}
H_{\mathbf S}(\omega)-H_{\mathbf A}(\omega)
 =\mathbf c^\top\mathbf R_{\mathbf S}(\mathbf S-\mathbf A)
  \bm\Gamma\mathbf R_{\mathbf A}\mathbf b,
\label{eq:response_difference}
\end{equation}
where $H_{\mathbf A}(\omega)$ and $H_{\mathbf S}(\omega)$ denote~\eqref{eq:fdn_transfer} at $z=e^{j\omega}$ using $\mathbf A$ and $\mathbf S$, respectively. Both use $d=0$ and the same delays, gains, and attenuation. Near a resonance, the resolvents can magnify a small feedback perturbation. The acoustic phase therefore minimises the log-spectral distance (LSD)
\begin{equation}
\LSD(\mathbf A,\mathbf S)=
\left[\frac{1}{K}\sum_{k=1}^{K}
\left(20\log_{10}\frac{|H_{\mathbf A}(\omega_k)|}
{|H_{\mathbf S}(\omega_k)|}\right)^2\right]^{1/2},
\label{eq:lsd}
\end{equation}
 evaluated over a uniform frequency grid $\omega_k=\pi(k-1)/K$, $K=512$. An offset of $10^{-30}$ is added to both magnitudes before taking logarithms. The responses are evaluated using batched complex linear solves. Optimisation is performed using Adam~\cite{kingma2015adam}, differentiated through PyTorch~\cite{paszke2019pytorch}, updating all coordinates with learning rate $5\times10^{-3}$ for 300--500 iterations at $\gamma=0.999$. The final block matrices are rescored in double precision. Unlike the exact block sweeps, individual Adam steps need not monotonically decrease the objective.

The depth $L$ was selected separately for each source and block size through informal engineering listening checks during development. Among the candidates compared, the lowest depth with no clearly audible differences from the dense reference was retained.
These checks provided a development stopping rule. The formal listening study in Section~\ref{sec:listening} evaluated candidates at $N\in\{16,32,64,128\}$.

\section{Real-Time Implementation}
\label{sec:implementation}

The independent blocks admit single-instruction, multiple-data (SIMD) evaluation, which essentially allows for one instruction to perform the same operation on several data elements at a time~\cite{flynn1966}. The runtime encodes fixed routing in the kernel and stores precomputed block coefficients. It requires no trigonometric functions, dense matrix materialisation, dynamic allocation, or re-orthogonalisation. The pairwise kernel reads the first and second halves of the state vector, evaluates independent rotations with NEON fused multiply-add instructions, and uses an interleaving store to create the perfect-shuffle order for the next layer.

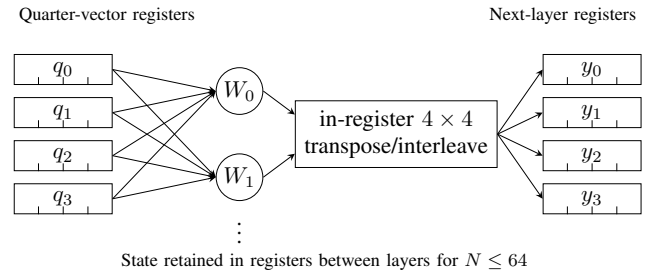
\begin{figure}[t]
    \centering
    \resizebox{0.98\columnwidth}{!}{\begin{tikzpicture}[>=stealth, line width=.4pt, font=\rmfamily\normalsize,
    reg/.style={draw, minimum width=1.5cm, minimum height=0.46cm, inner sep=1pt},
    blk/.style={draw, circle, inner sep=1.5pt, minimum size=0.6cm, fill=white},
    op/.style={draw, rectangle, align=center, inner sep=4pt}]

  \node[font=\small, anchor=south west] at (-0.8, 0.58) {Quarter-vector registers};
  \foreach \i/\y in {0/0, 1/-0.66, 2/-1.32, 3/-1.98} {
      \node[reg] (q\i) at (0, \y) {$q_{\i}$};
      \foreach \f in {0.25,0.5,0.75}
          \draw ($(q\i.south west)!\f!(q\i.south east)$) -- +(0,0.12);
  }

  \node[blk] (W0) at (2.7, -0.33) {$W_0$};
  \node[blk] (W1) at (2.7, -1.65) {$W_1$};
  \node at (2.7, -2.4) {$\vdots$};

  \foreach \i in {0,1,2,3} {
      \draw[->] (q\i.east) -- (W0.west);
      \draw[->] (q\i.east) -- (W1.west);
  }

  \node[op] (T) at (5.1, -0.99) {in-register $4\times4$\\ transpose/interleave};
  \draw[->] (W0.east) -- ($(T.west)+(0,0.3)$);
  \draw[->] (W1.east) -- ($(T.west)+(0,-0.3)$);

  \node[font=\small, anchor=south east] at (8.9, 0.58) {Next-layer registers};
  \foreach \i/\y in {0/0, 1/-0.66, 2/-1.32, 3/-1.98} {
      \node[reg] (y\i) at (8.1, \y) {$y_{\i}$};
      \foreach \f in {0.25,0.5,0.75}
          \draw ($(y\i.south west)!\f!(y\i.south east)$) -- +(0,0.12);
      \draw[->] (T.east) -- (y\i.west);
  }

  \node[font=\small, align=center] at (4.0, -2.95)
      {State retained in registers between layers for $N\leq64$};

\end{tikzpicture}}
    \caption{Register-resident radix-4 stage. Fixed quarter-vector gathers and the output interleave are expressed by register selection rather than stack traffic.}
    \label{fig:kernel}
\end{figure}

The radix-4 kernel follows the same principle one radix higher. Each layer gathers corresponding entries from four quarters of the state vector and applies one $4\times4$ block per output group. For $N\leq64$, the entire state occupies at most sixteen NEON registers. Specialised kernels keep it there through all layers, so the gather becomes register naming, the interleaved write becomes an in-register $4\times4$ transpose, and each dot product is tree-reduced. Figure~\ref{fig:kernel} illustrates this register layout and the transition between layers. Relative to the generic path with identical weights, this gives 1.7$\times$, 1.5$\times$, and 1.3$\times$ at $N=16,32,64$.

Table~\ref{tab:roofline} compares three block sizes at $N=256$, each with full structural spread, fitted to a random orthogonal target. The structured coefficient arrays occupy 8--32 KiB. Increasing the block size raises achieved arithmetic throughput, but also increases the work per multiply. Radix-4 balances these effects and reaches the lowest time. The measured rates explain why operation count alone does not predict execution time.

\begin{table}[t]
\centering
\caption{Block-size comparison at $N=256$ on an Apple M2. FLOPs follow the counting conventions of Section~III; the rates use these counts and the listed times.}
\label{tab:roofline}
\footnotesize
\setlength{\tabcolsep}{4pt}
\begin{tabular}{@{}lrrrr@{}}
\toprule
Structure & Nominal FLOPs & ns & GFLOP/s & $\tr(\mathbf A^\top\mathbf S)/N$ \\
\midrule
$b=2,L=8$  & 6144   & 202.5 & 30.3 & 0.219 \\
$b=4,L=4$  & 8192   & 152.9 & 53.6 & 0.270 \\
$b=16,L=2$ & 16384  & 262.6 & 62.4 & 0.392 \\
\bottomrule
\end{tabular}
\end{table}

Timings use FP32 kernels on an Apple M2 and column-major Apple Accelerate SGEMV. The harness times blocks of 512 calls and divides by the call count, while an in-loop memory clobber prevents fixed-input calls from being hoisted. The primary speed-up compares kernels and SGEMV under this same blocked protocol. Isolated SGEMV calls are also timed, including clock-reading overhead, and give a larger comparison ratio. Table~\ref{tab:headline} distinguishes these protocols; neither measures a complete FDN callback.

The pairwise Python convention is $\mathbf S_{\rm py}=(\mathbf P\mathbf B_{L-1})\cdots(\mathbf P\mathbf B_0)$, whereas the kernel applies $\mathbf B_\ell\mathbf P$, so  for a repeated permutation, $\mathbf S_{\rm kernel}=\mathbf P^\top\mathbf S_{\rm py}\mathbf P$. Export therefore relabels the delay states, delays, gains, and attenuation by $\mathbf P^\top$, which preserves the complete transfer function and adds no arithmetic in the callback. A heterogeneous routing schedule cannot be exported to this kernel unchanged.

\section{Evaluation}
\label{sec:evaluation}
We evaluate runtime and storage, magnitude-response agreement, modal excitation, echo density, and perceived similarity of the obtained staged approximations. Seven dense source systems, $N\in\{4,8,16,32,64,128,256\}$, act as the projection targets. A polar factor removes residual loss of orthogonality in each stored training matrix. Response fitting holds the trained delays and gains fixed and uses $\gamma=0.999$ with $K =512$ bins. The listening test uses separately rendered reverberated drum loops.

\subsection{Runtime and Storage}
Table~\ref{tab:headline} compares the cost of applying the pairwise and radix-4 structures in terms of feedback-multiply time and coefficient storage on Apple M2. Asterisks mark the selected operating points. At $N=4$, the implementation uses a direct $4\times4$ multiply that represents the dense reference exactly; this is an implementation choice, not a restriction of the staged family. At $N=8$, only the pairwise kernel is deployed, since the specialised radix-4 kernels begin at $N=16$.

Table~\ref{tab:headline} shows that for $N \in \{ 16,32,64\}$, radix-4 results in lower depth $L$ and is faster than the corresponding pairwise design. The largest difference occurs at $N=32$, where implementing four radix-4 layers takes 18.9 ns versus 55.4 ns for ten pairwise layers, at the same time increasing coefficient storage from 320 to 512 floats.

\begin{table}[t!]
\centering
\caption{Feedback-multiply time and coefficient storage on Apple M2. Asterisks mark selected implementations. Speedups use blocked and isolated dense SGEMV timings as defined in Section~\ref{sec:implementation}. The $N=128,256$ radix-4 rows report implementation cost only.}
\label{tab:headline}
\footnotesize
\setlength{\tabcolsep}{4pt}
\begin{tabular}{@{\raggedleft\arraybackslash}m{0.45cm}>{\raggedleft\arraybackslash}m{2.2cm}r>{\raggedleft\arraybackslash}m{1cm}>{\raggedleft\arraybackslash}m{1cm}>{\raggedleft\arraybackslash}m{1cm}>{\raggedleft\arraybackslash}m{1cm}@{}}
\toprule
$N$ & Structure & $L$ & Kernel (ns) & Blocked & Isolated & Stored floats \\
\midrule
4   & dense $4\times4$ exact$^\ast$ & 1 & 1.0 & $\sim30\times$ & --- & 16 \\
8   & Pairwise$^\ast$ & 6 & 11.3 & $\sim2.7\times$ & --- & 48 \\
\midrule
\multirow{2}{12pt}{16}  & Pairwise & 4 & 11.2 & 3.5$\times$ & 3.7$\times$ & 64 \\
 & Radix-4$^\ast$ & 2 & 4.5 & 8.7$\times$ & 9.4$\times$ & 128 \\
\multirow{2}{12pt}{32}   & Pairwise & 10 & 55.4 & 1.3$\times$ & 1.5$\times$ & 320 \\
  & Radix-4$^\ast$ & 4 & 18.9 & 3.8$\times$ & 4.4$\times$ & 512 \\
\multirow{2}{12pt}{64}   & Pairwise & 3 & 26.4 & 8.5$\times$ & 9.5$\times$ & 192 \\
  & Radix-4$^\ast$ & 2 & 18.2 & 12.3$\times$ & 13.7$\times$ & 512 \\
\midrule
\multirow{2}{12pt}{128}  & Pairwise$^\ast$ & 3 & 44.4 & 9.2$\times$ & 14.1$\times$ & 384 \\
 & Radix-4 & 2 & 42.0 & 9.7$\times$ & 14.9$\times$ & 1024 \\
\multirow{2}{12pt}{256}  & Pairwise$^\ast$ & 3 & 84.4 & 16.4$\times$ & 21.7$\times$ & 768 \\
 & Radix-4 & 2 & 86.4 & 16.0$\times$ & 21.2$\times$ & 2048 \\
\bottomrule
\end{tabular}
\end{table}

For larger systems, the pairwise structure uses substantially fewer coefficients with little difference in runtime. At $N=128$, it takes 44.4 ns and stores 384 coefficients, compared to 42.0 ns and 1024 coefficients for radix-4. At $N=256$, pairwise is also slightly faster, taking 84.4 ns and storing 768 coefficients, against 86.4 ns and 2048 coefficients for radix-4. These comparisons favour the pairwise implementation on storage grounds while their transfer function fits are considered separately in the following section.

The selected structures reduce coefficient storage by $8\times$ at $N=64$, $42.7\times$ at $N=128$, and $85.3\times$ at $N=256$, relative to the $N^2$ dense coefficients. In general, the reduction is $N/L$ for pairwise rotations and $N/(4L)$ for radix-4.

\subsection{Transfer Function Fitting}

Table~\ref{tab:depth_ablation} reports the LSD between the frequency response of each staged FDN and its corresponding dense reference.
At $N\in\{16,32,64\}$, the selected radix-4 fits have lower LSD than their pairwise counterparts. For example, at $N=32$, the error falls from 0.100 dB with ten pairwise layers to 0.078 dB with four radix-4 layers. Within the selected radix-4 structures, response refinement reduces LSD by 45--85\% without adding layers or coefficients. Further layers reduce the recorded errors at $N\in\{16,64\}$, although the shallower designs had already been chosen for further evaluation during development listening checks (\emph{cf.}~Sec.~\ref{sec:projection}). Conversely, the $N=32$ three-layer candidate was rejected in those checks despite its low sampled error, illustrating why LSD alone did not determine the chosen depth.

\begin{table}[!t]
\centering
\caption{Response-fit errors on the 512-bin fitting grid at $\gamma=0.999$. (a) Pairwise fits. (b) Radix-4 fits before and after acoustic refinement, with selected depths marked by asterisks. Delays, gains, and attenuation remain fixed. LSD measures sampled magnitude-response error, not perceptual similarity.}
\label{tab:depth_ablation}
\footnotesize
\setlength{\tabcolsep}{4pt}
\begin{tabular}{@{}lrrrrrr@{}}
\toprule
\multicolumn{7}{@{}l}{(a) Pairwise fits}\\
$N$ & 8 & 16 & 32 & 64 & 128 & 256\\
$L$ & 6 & 4 & 10 & 3 & 3 & 3\\
\midrule
LSD (dB) & 0.38 & 2.13 & 0.100 & 3.27 & 2.26 & 1.60\\
\bottomrule
\end{tabular}\\[4pt]
\begin{tabular}{@{}rrrrr@{}}
\toprule
\multicolumn{5}{@{}l}{(b) Radix-4 refinement}\\
$N$ & $L$ & Before (dB) & After (dB) & Reduction \\
\midrule
16 & $2^\ast$ & 3.122 & 1.719 & 45\% \\
16 & 3 & 2.237 & 0.838 & 63\% \\
32 & 2 & 0.691 & 0.259 & 62\% \\
32 & 3 & 0.632 & 0.170 & 73\% \\
32 & $4^\ast$ & 0.509 & 0.078 & 85\% \\
64 & $2^\ast$ & 5.542 & 2.177 & 61\% \\
64 & 3 & 5.064 & 1.256 & 75\% \\
\bottomrule
\end{tabular}
\end{table}

\subsection{Established Acoustic Measures}

Magnitude-response agreement does not fully describe the temporal development or modal excitation of a reverberator. We therefore complement the response-fitting objective with the standard deviation of modal excitation, as a measure of spectral unevenness \cite{dalsanto2023, heldmann2021}, and a mixing-time estimate derived from echo density \cite{abel2006, schlecht2017echo}.

For distinct poles, the impulse response at $n\geq1$ has the modal expansion
\begin{equation}
 h(n)=\sum_{i=1}^{\mathfrak N} h_i(n),\qquad
 h_i(n)=|\rho_i|\,|\lambda_i|^n
 e^{j(n\angle\lambda_i+\angle\rho_i)},
\label{eq:modal_decomposition}
\end{equation}
where $\mathfrak N=\sum_i m_i$ is the system order and $\rho_i$ is the coefficient of the corresponding one-pole term $(1-\lambda_i z^{-1})^{-1}$. The initial sample is handled separately, with $h(0)=d=0$ for the evaluated systems.
The poles $\lambda_i$ determine frequency and decay rate of each mode $h_i(n)$, while residues $\rho_i$ determine the modal excitation and initial phase~\cite{schlecht2019modal}. Following~\cite{heldmann2021}, we measure the spread of modal excitation by
\begin{equation}
 \sigma_{\rho}=\operatorname{std}\!\left(
 20\log_{10}\frac{|\rho_i|}{\max_j|\rho_j|}\right).
\label{eq:modal_spread}
\end{equation}
A smaller $\sigma_{\rho}$ generally indicates a more even excitation distribution and less colour in the late decay \cite{dalsanto2023, heldmann2021}. However, this metric is not sufficient by itself---a reverberator may have acceptable modal properties while producing sparse output with audible discrete echoes \cite{dalsanto2023, dalsanto2025tiny}. We estimate mixing time $t_{\mathrm{mix}}$ from Abel--Huang normalised echo density~\cite{abel2006,schlecht2017echo}. In each non-overlapping 10\,ms window, the fraction of samples whose magnitude exceeds the window RMS is divided by the Gaussian reference value 0.3173. The centre of the first window reaching 0.95 defines $t_{\mathrm{mix}}$. This is an echo-density threshold, not a direct perceptual measurement.

Table~\ref{tab:acoustic} reports the original acoustic measurements for dense references and their pairwise projections. Its $N=4$ result uses an eight-layer fit; the runtime table instead uses an exact direct multiply. Full modal decomposition was not computed for $N\geq128$ because of the large system order $\mathfrak N$. Modal excitation spread remains unchanged at the printed precision for three of the five measured FDN sizes. Its largest change is 0.7 dB, at $N=32$. Mixing-time differences range from $-20$ to $+50$ ms. The pairwise projections thus approximately preserve these measures for the tested systems. Repeating these checks for the selected radix-4 projections at $N\in\{16,32,64\}$, with dense values recomputed under matched protocols, gives modal-excitation-spread changes below 0.7\,dB and mixing-time increases of 30--70\,ms on the trained-gain renders. The radix-4 projections therefore preserve these measures to a similar degree as the pairwise fits.

\begin{table}[t]
\centering
\caption{Acoustic measures for the original pairwise projections. Dashes indicate values not computed.}
\label{tab:acoustic}
\footnotesize
\setlength{\tabcolsep}{4pt}
\begin{tabular}{@{}l lrrrrrrr@{}}
\toprule
& $N$ & 4 & 8 & 16 & 32 & 64& 128 & 256\\
& $L$ & 8 & 6 & 4 & 10 & 3 & 3 & 3\\
\midrule
\multirow{2}{30pt}{ $\sigma_\rho$ (dB)} & Dense & 3.0 & 6.8 & 8.2 & 7.1 & 7.8 & - & -\\
& Staged & 3.0 & 6.8 & 8.0 & 7.8 & 7.8 & - & -\\
\midrule
\multirow{2}{36pt}{ $t_{\rm mix}$ (ms)} & Dense & 1325 & 205 & 125 & 305 & 85 & 95 & 75\\
& Staged & 1325 & 205 & 175 & 305 & 65 & 95 & 115\\
\bottomrule
\end{tabular}\\[4pt]

\end{table}

\begin{figure*}[t]
    \centering
    \includegraphics[width=\textwidth]{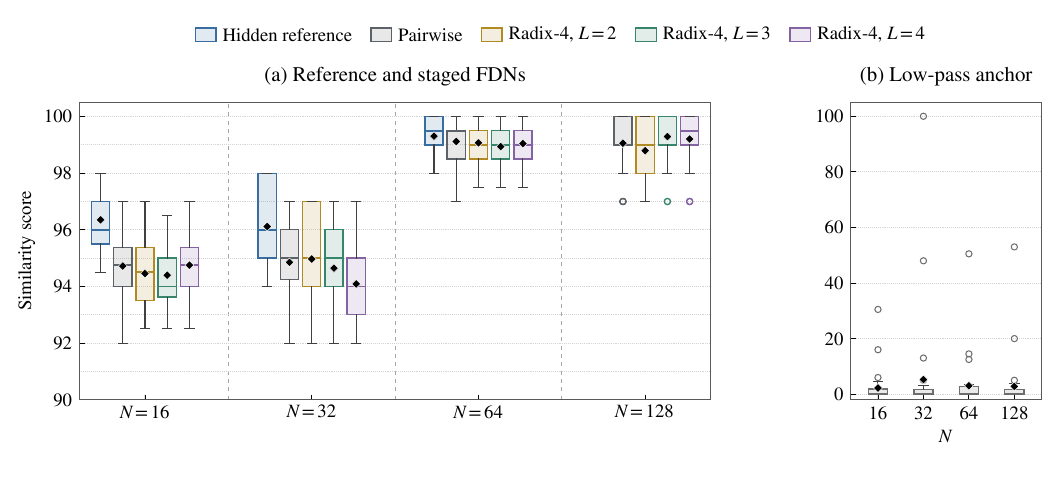}
    \caption{Listening-test similarity scores ($n=34$). (a) Hidden references and staged FDNs, enlarged to the 90--100 range. (b) Low-pass anchors on the full 0--100 scale. Boxes span the interquartile range, horizontal lines mark medians, and black diamonds mark means. Whiskers extend to observations within $1.5$ interquartile ranges of the box; circles show outliers. Repeated ratings are averaged within each participant. Pairwise depths are $L=4,10,3,3$ for increasing $N$.}
    \label{fig:listening}
\end{figure*}

\subsection{Listening Study}\label{sec:listening}

We assessed the perceived similarity between staged and dense FDN reverberation in an online MUSHRA-style experiment~\cite{itu2015mushra}, implemented with goListen~\cite{golisten}. Listeners compared versions of a reverberated drum loop against an explicit dense reference and rated similarity from 0 to 100, with 100 labelled ``indistinguishable from the reference.''

The experiment covered $N\in\{16,32,64,128\}$. At each size, the candidates comprised one pairwise projection and radix-4 projections with $L\in\{2,3,4\}$, together with a low-pass anchor. The pairwise depths were $L=4,10,3,3$, respectively. Unintended test-setup omissions left two presentations each of $N=16,64$, one each of $N=32,128$, and no scored hidden reference at $N=128$. Page and candidate order were randomised. The explicit dense reference was available on every page; hidden references, where present, used the same audio.

The study included 34 consenting participants using their own headphones or earphones, with instructions to complete the test undistracted in a quiet room. Participants were recruited primarily through the authors' institutions and collaborating audio research groups. A questionnaire recorded gender, audio experience, self-reported hearing, equipment, and listening environment; participant age was not recorded. Of the analysed participants, 22 identified as men, 11 as women, and one as non-binary. Most participants were researchers or students in audio or acoustics, and the remainder were musicians or audio enthusiasts. No participant reported a hearing impairment. During practice, listeners were asked to set a comfortable playback level and retain it throughout the test. All 34 passed the BS.1534 hidden-reference screen on the available reference ratings~\cite{itu2015mushra}.

Repeated ratings were averaged within each participant before calculating group means and sample standard deviations, giving every configuration the same $n=34$. Figure~\ref{fig:listening} separates the reference and candidate distributions from the low-pass anchors as the large discrepancy of scores would otherwise hinder the readability of results. Anchors' low scores indicate that listeners distinguished the deliberately degraded signal. Numerical summaries and the screening rule are provided in the supplementary analysis.

The mean scores of the staged approximation candidates ranged from 94.09 to 94.97 at $N=16,32$ and from 98.79 to 99.29 at $N=64,128$, with between-participant standard deviations of 0.64--1.80 points. The selected radix-4 candidates scored 94.46, 94.09, and 99.07 at $N=16,32,64$; the selected pairwise $N=128,L=3$ candidate scored 99.06. Anchor means spanned 2.22--5.26. Candidate means lay approximately one to two points below the hidden references at $N=16,32$, while both approached the top of the scale at $N=64$.

The tested configurations were thus rated close to their dense references, with the higher-order candidates near the scale's ``indistinguishable'' endpoint. Additional layers did not consistently improve the ratings. The study concerns one excerpt and a different dense reference at each order, so it does not isolate an effect of network size or establish perceptual equivalence.

\subsection{Numerical Verification}
\label{sec:numerics}

Numerical checks distinguish exact orthogonality from its finite-precision evaluation~\cite{trefethen1997,higham1996accuracy}. For the tested random products, double-precision residuals $\|\mathbf S^\top\mathbf S-\mathbf I\|_F$ and structured-versus-dense application errors were below $10^{-12}$. The FP32 radix-4 kernels met $10^{-4}$ tolerances on maximum output error and relative energy error against a scalar reference with double accumulation, across fourteen configurations and 32 seeded vectors each. Separate checks verified pairwise export and transfer-function invariance under channel relabelling. The accompanying code includes recovery tests, response-refinement checks, and numerical logs.

\section{Discussion}
\label{sec:discussion}

For the routing used here, the connectivity bound identifies the depth at which every input first has a structural path to every output. It describes paths through one feedback multiplication, rather than the elapsed mixing time of the reverberator. The routing matters here as much as the layer count. With fixed permutations, the families obtained at successive depths need not be nested, so a deeper structure need not achieve a smaller error. Exact block updates solve the conditional problems, but the joint fit remains a local optimisation. The complete FDN response therefore provides the final criterion for choosing depth.

Block size determines how this approximation is executed. The radix-4 results show why a useful comparison must include data movement and register reuse alongside arithmetic. Larger blocks expose more independent operations to the processor, but their additional work can outweigh that benefit, as the block-size comparison illustrates. Measured execution time, response error, and coefficient storage together give a practical basis for selecting a structure. The present timings cover feedback multiplication on an Apple M2; the speedup of a complete reverberator also depends on its delay access, attenuation filters, and accumulation.

These results suggest studying routing and block size together, including mixed blocks and implementations on other processors. A bound relating response error to $N$, $b$, and $L$ would further guide this choice. The projection code, source matrices, kernels, benchmark harness, and full listening analysis are intended for release with the article through the project repository.\footnote{\url{https://github.com/cucuwritescode/fdn_pruning}}

\raggedbottom
\section{Conclusion}
\label{sec:conclusion}

This work proposes a new paradigm in FDN design. Rather than constraining the structure first and recovering its sound afterwards, we begin with the desired acoustic response and make its implementation substantially cheaper.
A dense FDN is first designed or selected for its reverberation, then its feedback matrix is fitted to a product of fixed permutations and small orthogonal blocks. We studied Givens-rotation stages ($b=2$) and radix-4 Pease stages ($b=4$), using exact conditional block updates for matrix fitting, followed by joint refinement against the complete FDN magnitude response. Orthogonality is preserved by construction, while delays, gains, and attenuation remain fixed to the learned target design.

The resulting kernels reduce feedback-multiplication time and coefficient storage on the tested Apple M2. Modal-excitation spread and mixing time are approximately preserved in the measured pairwise and radix-4 cases, and listeners rate the tested higher-order configurations near the scale's indistinguishable endpoint. Block size and depth can therefore be chosen to reduce implementation cost after the desired acoustic behaviour has been established. The resulting approximation is assessed in the complete reverberator, offering a practical route towards the ideal FDN triad of colourlessness, smoothness, and low computational cost.

\Needspace*{14\baselineskip}
\section*{ACKNOWLEDGMENT}
The authors thank Prof. Sebastian J. Schlecht for his valuable advice in refining the research question, and Jatin Chowdhury for his ideas and resources on SIMD implementation.

\bibliographystyle{IEEEbib_final_rev}
\bibliography{refs_final_rev}

\begin{thebibliography}{10}

\bibitem{jot1991}
J.-M. Jot and A.~Chaigne,
\newblock ``Digital delay networks for designing artificial reverberators,''
\newblock in {\em Proceedings of the 90th AES Convention}, 1991.

\bibitem{stautner1982multichannel}
J.~Stautner and M.~Puckette,
\newblock ``Designing multi-channel reverberators,''
\newblock {\em Computer Music Journal}, vol. 6, no. 1, pp. 52--65, 1982.

\bibitem{valimaki2012fifty}
V.~V{\"a}lim{\"a}ki, J.~D. Parker, L.~Savioja, J.~O. Smith, and J.~S. Abel,
\newblock ``Fifty years of artificial reverberation,''
\newblock {\em IEEE Transactions on Audio, Speech, and Language Processing},
  vol. 20, no. 5, pp. 1421--1448, 2012.

\bibitem{jot1997efficient}
J.-M. Jot,
\newblock ``Efficient models for reverberation and distance rendering in
  computer music and virtual audio reality,''
\newblock in {\em Proc. ICMC}, 1997.

\bibitem{menzer2010unitary}
F.~Menzer and C.~Faller,
\newblock ``Unitary matrix design for diffuse {J}ot reverberators,''
\newblock in {\em Proceedings of the 128th Audio Engineering Society
  Convention}, London, UK, 2010,
\newblock Convention paper 7984.

\bibitem{schlecht2020fdntb}
S.~J. Schlecht,
\newblock ``{FDNTB}: The feedback delay network toolbox,''
\newblock in {\em Proc. Int. Conf. Digit. Audio Effects}, 2020, pp. 211--218.

\bibitem{dalsanto2023}
G.~Dal~Santo, K.~Prawda, S.~J. Schlecht, and V.~V{\"a}lim{\"a}ki,
\newblock ``Differentiable feedback delay network for colorless
  reverberation,''
\newblock in {\em Proceedings of the International Conference on Digital Audio
  Effects (DAFx)}, 2023.

\bibitem{dalsanto2025tiny}
G.~Dal~Santo, K.~Prawda, S.~J. Schlecht, and V.~V{\"a}lim{\"a}ki,
\newblock ``Optimizing tiny colorless feedback delay networks,''
\newblock {\em EURASIP Journal on Audio, Speech, and Music Processing}, vol.
  2025, no. 1, pp. 13, 2025.

\bibitem{schlecht2017lossless}
S.~J. Schlecht and E.~A.~P. Habets,
\newblock ``On lossless feedback delay networks,''
\newblock {\em IEEE Transactions on Signal Processing}, vol. 65, no. 6, pp.
  1554--1564, 2017.

\bibitem{schlecht2017echo}
S.~J. Schlecht and E.~A.~P. Habets,
\newblock ``Feedback delay networks: Echo density and mixing time,''
\newblock {\em IEEE/ACM Transactions on Audio, Speech, and Language
  Processing}, vol. 25, no. 2, pp. 374--383, 2017.

\bibitem{santo2025}
G.~Dal~Santo, G.~M. De~Bortoli, K.~Prawda, S.~J. Schlecht, and
  V.~V{\"a}lim{\"a}ki,
\newblock ``{FLAMO}: An open-source library for frequency-domain differentiable
  audio processing,'' 2025.

\bibitem{gerzon1976}
M.~A. Gerzon,
\newblock ``Unitary energy-preserving multichannel networks with feedback,''
\newblock {\em Electronics Letters}, vol. 12, no. 11, pp. 278--279, 1976.

\bibitem{householder1958triangularization}
A.~S. Householder,
\newblock ``Unitary triangularization of a nonsymmetric matrix,''
\newblock {\em Journal of the ACM}, vol. 5, no. 4, pp. 339--342, 1958.

\bibitem{hadamard1893resolution}
J.~Hadamard,
\newblock ``R{\'e}solution d'une question relative aux d{\'e}terminants,''
\newblock {\em Bulletin des Sciences Math{\'e}matiques}, vol. 17, pp. 240--246,
  1893.

\bibitem{rocchesso1997}
D.~Rocchesso and J.~O. Smith,
\newblock ``Circulant and elliptic feedback delay networks for artificial
  reverberation,''
\newblock {\em IEEE Transactions on Speech and Audio Processing}, vol. 5, no.
  1, pp. 51--63, 1997.

\bibitem{heldmann2021}
J.~Heldmann and S.~J. Schlecht,
\newblock ``The role of modal excitation in colorless reverberation,''
\newblock in {\em Proceedings of the 24th International Conference on Digital
  Audio Effects (DAFx)}, Vienna, Austria, 2021, pp. 206--213.

\bibitem{prawda2020flexible}
K.~Prawda, S.~Willemsen, S.~Serafin, and V.~V{\"a}lim{\"a}ki,
\newblock ``Flexible real-time reverberation synthesis with accurate parameter
  control,''
\newblock in {\em Proc. DAFx}, 2020.

\bibitem{anderson2015flatter}
H.~Anderson, K.~W.~E. Lin, C.~So, and S.~Lui,
\newblock ``Flatter frequency response from feedback delay network reverbs,''
\newblock in {\em Proc. ICMC}, 2015.

\bibitem{schlecht2026pyfdntoolbox}
S.~J. Schlecht, F.~Franchino, and J.~B. Bai,
\newblock ``{pyFDN}: A python toolbox for feedback delay network design,
  analysis, and optimization,''
\newblock {\em arXiv preprint arXiv:2610.27217}, 2026.

\bibitem{franchino2026}
F.~Franchino and S.~J. Schlecht,
\newblock ``Compiling differentiable audio graphs to real-time {DSP},''
\newblock in {\em Proc. Int. Conf. Digital Audio Effects (DAFx26)}, 2026,
\newblock Demonstrations track.

\bibitem{gerzon1971synthetic1}
M.~A. Gerzon,
\newblock ``Synthetic stereo reverberation,''
\newblock {\em Studio Sound}, vol. 13, no. 12, pp. 632--635, Dec. 1971.

\bibitem{coppola2026fast}
A.~Coppola,
\newblock ``Fast parametric matrices for lossless feedback delay networks,''
\newblock in {\em Proceedings of the 29th International Conference on Digital
  Audio Effects (DAFx26)}, Cambridge, UK, 2026.

\bibitem{frankle2019lottery}
J.~Frankle and M.~Carbin,
\newblock ``The lottery ticket hypothesis: Finding sparse, trainable neural
  networks,''
\newblock in {\em Proc. Int. Conf. Learning Representations (ICLR)}, 2019.

\bibitem{clarke2025}
C.~J. Clarke and J.~Chowdhury,
\newblock ``Inference-time structured pruning for real-time neural network
  audio effects,''
\newblock in {\em Proc. 28th Int. Conf. Digital Audio Effects (DAFx25)},
  Ancona, Italy, 2025, pp. 358--365.

\bibitem{duff2017}
I.~S. Duff, A.~M. Erisman, and J.~K. Reid,
\newblock {\em Direct Methods for Sparse Matrices},
\newblock Numerical Methods and Scientific Computation. Oxford University
  Press, 2nd edition, 2017.

\bibitem{cantero2003cmv}
M.~J. Cantero, L.~Moral, and L.~Vel{\'a}zquez,
\newblock ``Five-diagonal matrices and zeros of orthogonal polynomials on the
  unit circle,''
\newblock {\em Linear Algebra and its Applications}, vol. 362, pp. 29--56,
  2003.

\bibitem{simon2005}
B.~Simon,
\newblock {\em Orthogonal Polynomials on the Unit Circle},
\newblock American Mathematical Society, 2005.

\bibitem{golub2013}
G.~H. Golub and C.~F. Van~Loan,
\newblock {\em Matrix Computations},
\newblock Johns Hopkins University Press, Baltimore, MD, 3rd edition, 1996.

\bibitem{givens1958rotations}
W.~Givens,
\newblock ``Computation of plane unitary rotations transforming a general
  matrix to triangular form,''
\newblock {\em Journal of the Society for Industrial and Applied Mathematics},
  vol. 6, no. 1, pp. 26--50, 1958.

\bibitem{diaconis1983}
P.~Diaconis, R.~L. Graham, and W.~M. Kantor,
\newblock ``The mathematics of perfect shuffles,''
\newblock {\em Advances in Applied Mathematics}, vol. 4, no. 2, pp. 175--196,
  1983.

\bibitem{pease1968}
M.~C. Pease,
\newblock ``An adaptation of the fast fourier transform for parallel
  processing,''
\newblock {\em Journal of the ACM}, vol. 15, no. 2, pp. 252--264, 1968.

\bibitem{schonemann1966}
P.~H. Sch{\"o}nemann,
\newblock ``A generalized solution of the orthogonal {Procrustes} problem,''
\newblock {\em Psychometrika}, vol. 31, no. 1, pp. 1--10, 1966.

\bibitem{cayley1846}
A.~Cayley,
\newblock ``Sur quelques propri\'{e}t\'{e}s des d\'{e}terminants gauches,''
\newblock {\em Journal f\"{u}r die reine und angewandte Mathematik}, vol. 32,
  pp. 119--123, 1846.

\bibitem{wen2013feasible}
Z.~Wen and W.~Yin,
\newblock ``A feasible method for optimization with orthogonality
  constraints,''
\newblock {\em Mathematical Programming}, vol. 142, pp. 397--434, 2013.

\bibitem{edelman1998}
A.~Edelman, T.~A. Arias, and S.~T. Smith,
\newblock ``The geometry of algorithms with orthogonality constraints,''
\newblock {\em SIAM Journal on Matrix Analysis and Applications}, vol. 20, no.
  2, pp. 303--353, 1998.

\bibitem{kingma2015adam}
D.~P. Kingma and J.~Ba,
\newblock ``Adam: A method for stochastic optimization,''
\newblock in {\em Proceedings of the International Conference on Learning
  Representations (ICLR)}, 2015.

\bibitem{paszke2019pytorch}
A.~Paszke, S.~Gross, F.~Massa, A.~Lerer, J.~Bradbury, G.~Chanan, T.~Killeen,
  Z.~Lin, N.~Gimelshein, L.~Antiga, A.~Desmaison, A.~K{\"o}pf, E.~Yang,
  Z.~DeVito, M.~Raison, A.~Tejani, S.~Chilamkurthy, B.~Steiner, L.~Fang,
  J.~Bai, and S.~Chintala,
\newblock ``{PyTorch}: An imperative style, high-performance deep learning
  library,''
\newblock in {\em Advances in Neural Information Processing Systems}, 2019,
  vol.~32.

\bibitem{flynn1966}
M.~J. Flynn,
\newblock ``Very high-speed computing systems,''
\newblock {\em Proceedings of the IEEE}, vol. 54, no. 12, pp. 1901--1909, 1966.

\bibitem{abel2006}
J.~S. Abel and P.~Huang,
\newblock ``A simple, robust measure of reverberation echo density,''
\newblock in {\em Proceedings of the 121st Audio Engineering Society
  Convention}, San Francisco, CA, USA, 2006,
\newblock Convention paper 6985.

\bibitem{schlecht2019modal}
S.~J. Schlecht and E.~A.~P. Habets,
\newblock ``Modal decomposition of feedback delay networks,''
\newblock {\em IEEE Transactions on Signal Processing}, vol. 67, no. 20, pp.
  5340--5351, 2019.

\bibitem{itu2015mushra}
{ITU-R},
\newblock ``Method for the subjective assessment of intermediate quality level
  of audio systems,''
\newblock Recommendation BS.1534-3, International Telecommunication Union,
  2015.

\bibitem{golisten}
{University College Dublin},
\newblock ``{goListen: A tool for perceptual listening tests},''
  \url{https://golisten.ucd.ie/}, 2026.

\bibitem{trefethen1997}
L.~N. Trefethen and D.~Bau, III,
\newblock {\em Numerical Linear Algebra},
\newblock Society for Industrial and Applied Mathematics, Philadelphia, PA,
  USA, 1997.

\bibitem{higham1996accuracy}
N.~J. Higham,
\newblock {\em Accuracy and Stability of Numerical Algorithms},
\newblock Society for Industrial and Applied Mathematics, Philadelphia, PA,
  1996.

\end{thebibliography}

\begin{IEEEbiography}[{\parbox[b][1.25in][t]{1in}{\includegraphics[width=1in]{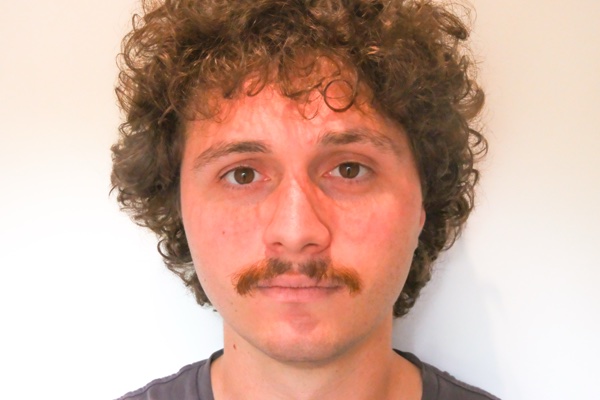}}}]
{Facundo Franchino}~is a graduate student at Massachusetts Institute of Technology (MIT), where he works under the supervision of Prof. Paris Smaragdis from the Research Laboratory of Electronics (RLE). He received the B.Eng. degree with first-class honours from the University of York, in 2026, where he was supervised by Dr. Karolina Prawda. His research interests include efficient signal processing and machine learning.
\end{IEEEbiography}

\begin{IEEEbiography}[{\parbox[b][1.25in][t]{1in}{\includegraphics[width=1in]{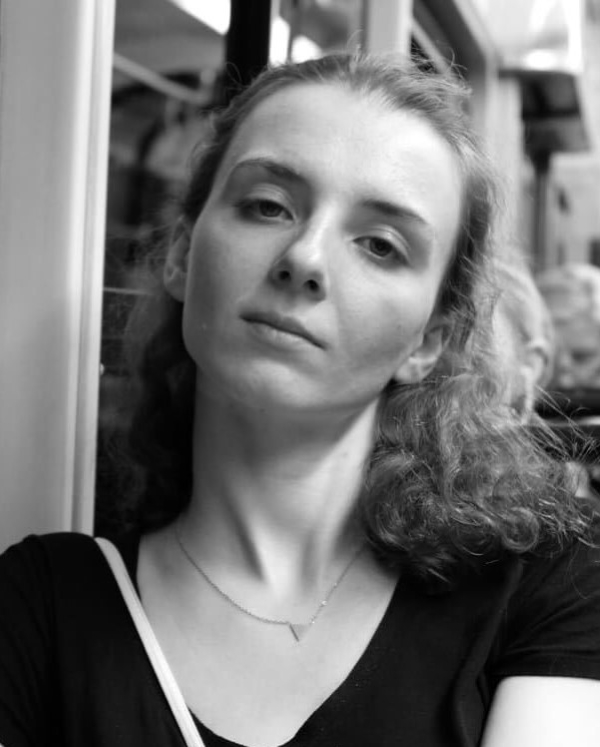}}}]
{Karolina Prawda}~is a Lecturer (Assistant Professor) in audio engineering for creative technologies at the AudioLab of the School of Physics, Engineering and Technology at the University of York. In 2017, she received a M.Sc. degree in acoustic engineering from AGH University of Kraków, Poland, and in 2022 she was awarded a D.Sc. (Tech.) degree in acoustics and audio signal processing from Aalto University, Finland. In 2022--2024, she was a Postdoctoral researcher at the Department of Information and Communications Engineering of Aalto University. Her research interests include acoustic metrology, room acoustic representation learning, and artificial reverberation. Dr.~Prawda is a member of the Audio Engineering Society. 
\end{IEEEbiography}

\end{document}